\documentclass[rnote]{aa}
\usepackage{graphicx}
\usepackage{txfonts}

\begin{document}

   \title{Quasar optical variability: searching for interband time delays}

   \author{Rumen S. Bachev}

   \offprints{Bachev}

   \institute{Institute of Astronomy, Bulgarian Academy of Sciences, 72 Tsarigradsko Chausse Blvd., 
               1784 Sofia, Bulgaria\\
              \email{bachevr@astro.bas.bg}}
   \date{Received ...; accepted ...}

\titlerunning{Quasar optical variability}
\authorrunning{Bachev}

\abstract
   {} 
   {The main purpose of this paper is to study time delays between the light variations in different wavebands for a sample of quasars.
Measuring a reliable time delay for a large number of quasars may help constraint the models of their central engines. The standard
accretion disk irradiation model predicts a delay of the longer wavelengths behind the shorter ones, a delay that depends on the fundamental quasar
parameters. Since the black hole masses and the accretion rates are approximately known for the sample we use, one can compare the observed
time delays with the expected ones.}
   {We applied the interpolation cross-correlation function ($ICCF$) method to the Giveon et al. sample of 42 quasars, monitored in two 
($B$ and $R$) colors, to find the time lags represented by the $ICCF$ peaks. Different tests were performed to assess the influence
of photometric errors, sampling, etc., on the final result.}
   {We found that most of the objects show a delay in the red light curve behind the blue one (a positive lag), 
which on average for the sample is about +4 days (+3 for the median), although the scatter is significant.
These results are broadly consistent with the reprocessing model, especially for the well-sampled objects. The normalized time-lag deviations 
do not seem to correlate significantly with other quasar properties, including optical, radio, or X-ray measurables. 
On the other hand, many objects show a clear negative lag, which, if real, may have important consequences for the variability models.}
   {} 


\maketitle

\section{Introduction}

Although variability is an extensively studied and rather common feature of quasars, little is currently known about the exact 
nature of the processes driving the flux changes. Variations are observed in practically all energy bands, often with 
time delays between them. This fact suggests a common cause (process) connecting otherwise spatially separated regions (presumably 
parts of an accretion disk), where most of the corresponding wavelengths come from. Since the time lags are rather short for 
the expected distances, it is the speed of light that is perhaps the only way to casually connect these separate regions. 
Several competing ideas, often leading to different signs of the time lags between the bands have been discussed 
in the literature (Czerny et al. 2008):

\begin{itemize} 
\item {\bf Reprocessing} of the central high-energy continuum into lower energy bands from the outer (and colder) regions of an 
accretion disk (Krolik et al. 1991). For this mechanism to work, the central X-ray emission must ``see'' the outer parts -- either 
because of elevation of the central source above the disk surface, of warping, or of flaring of the disk (or all three). Thus, a time 
lag between the variable X-rays and the lower wavelength (e.g. UV, optical, IR) will be observed with a relation between the lag and the 
wavelength, roughly scaled as $\tau_{\rm \lambda} \sim \lambda^{4/3}$ (Sect. 4) for a standard accretion disk (Shakura \& Syunyaev 1973). 
Therefore, if the reprocessing is responsible for the most of the optical variability, a delay between every two optical bands 
(say $B$ and $R$-bands) will be expected, with the shorter wavelengths leading (a positive lag). 
However, if the variable X-rays come not from the center, but from another location instead (e.g. from local flares above the disk surface), 
the sign of the lags can be either positive or negative, depending on the radial distance at which these flares predominantly occur.
In this case, a connection between the signs of the lags and the fundamental quasar parameters could be searched, as the radial
distance of the X-ray flares is probably governed by the structure of a corona above the disk, which in turn should depend on the
quasar parameters.

\item {\bf A weak blazar component}, often assumed to exist even in radio-quiet sources (Czerny et al. 2008, and the references therein). 
Such a component can produce variable optical/IR synchrotron radiation, and respectively -- variable X-rays via the synchrotron 
self-Compton mechanism. These X-rays can, in turn, be reprocessed in a disk into optical/UV photons, leading to a complicated interband 
time-delay picture, in which the redder (synchrotron) variations will probably lead the bluer (reprocessed) ones.


\item {\bf Disk fluctuations}, drifting inward. This mechanism is sometimes invoked (Ar\'evalo et al. 2008) to explain cases 
where the long-wavelength variation are leading. However, the viscous timescale at the optical band generation distance is much 
longer than the timescales of a few days, discussed in this paper.

\end{itemize} 

The situation can be very complicated, as several of the effects mentioned above can work in combination, or of course, 
an entirely different mechanism can be responsible for the interband relations. In either case, studying the lags between the 
continuum changes in different wavelengths for a large sample of quasars may help to clarify the situation, and therefore, 
to create a better picture of the quasar central engines. 

Sergeev et al. (2005) studied the light curves of 14 nearby Seyfert galaxies, observed on $\sim$60--150 epochs in 4
broad-band filters and found positive delays for most of the cases, varying between a fraction of a day and a few days. Their
results are broadly consistent with the reprocessing model. Recently, similar results were reported by Liu et al. (2008) and
Ar\'evalo et al. (2008). Here we extend their works, applying a similar technique to a larger 
sample of quasars, optically monitored at the Wise observatory (Giveon et al. 1999) in two colors for several years. 
The sample is described in more detail in the next section. In Sect. 3 we describe the method we used to find the 
time delay between the bands, i.e. the interpolation cross-correlation function ($ICCF$) method. Next, the results for the time lags 
are presented and compared with the predictions of the reprocessing model.
Finally, we discuss the reliability of the lag estimates and different implications for the central engine models of quasars,
by comparing the deviations from the reprocessing model predictions with various quasar characteristics.

\section{The sample}

This work analyses publicly available light curve data for a sample of 42 Palomar-Green (Schmidt \& Green 1983), 
mostly radio-quiet quasars, monitored at the Wise observatory in two observers-frame colors ($B$ and $R$-bands), Giveon et al. (1999). 
This sample was chosen mostly because of the high accuracy of the CCD photometry, typically around 0.01 mag, 
in comparison with earlier, photographic plate based monitoring campaigns (Giveon et al., 1999, and the references therein). 
The time span of the monitoring was about 7 year, with a typical average monitoring interval of $\sim40$ days, 
though ranging significantly. The best sampled objects were observed on about 80 separate epochs, while the least sampled -- on only 
about 25 epochs (see Sect. 5.1.1 for a discussion on sampling issues). All the objects are nearby, typically of $z\simeq0.2$, which insures 
that the analysis applies to almost the same rest-frame wavelength region. As an additional argument for using this sample is that
many of the objects have reverberation-mapping central mass estimates (and respectively -- Eddington ratios), so one can study
possible relations between the time lags and the accretion parameters. For the rest of the objects, due to the good optical spectra
available, the ``size--luminosity'' (Kaspi et al. 2005) relation can be applied for the same purpose.
The sample is presented in Table 1. The object name, the red shift and the number of observational epochs are
given in the first three columns, followed by the measured time lags between $B$ and $R$-bands, $\tau_{\rm obs}$ (see Sect. 4 for details).
Next columns show the black hole mass, the accretion rate (in Eddington units) and the expected lag for the simple reprocessing model (Sect. 4).

\section{The $ICCF$ method}

In order to study the wavelength -- time delay dependence we performed a linear-interpolation cross-correlation ($ICCF$)
analysis (Gaskell \& Sparke 1986) between the light curves of the two bands. The maximum of $ICCF(\tau)$ is assumed to give the time
delay between the bands. The interpolation between the photometric points is necessary due to the unevenly sampled data and is one of 
the frequently used methods. In our particular case, the magnitude values for every second day of the interpolated light curve were 
later on used for the cross-correlation analysis. This time interval of 2 days seems like a reasonable choice, since it is shorter than the
majority of the real data intervals, but is not too short so a huge number of ``artificial'' data points to influence the
analysis. Such a 2-day interval leads to an additional ($rms$) uncertainty of $\sigma_{\rm int}\simeq0.6$ days in the peak position, which 
however is generally smaller than the expected errors of other nature (see Sect. 5.1).  

Other methods described in the literature (e.g. discrete $CCF$, Edelson \& Krolik 1988; $z$-transformed $CCF$, Alexander 1997) do not 
seem to give significantly different results when applied to the same data (White \& Peterson 1994; see also Fig. 3).
The weighted delays of the top 80\% of the $ICCF$ are typically 1.5 -- 2 times larger for this sample (see also Sergeev et al., 2005, 
for similar results). They correlate significantly (0.93) with the peak values, which are used throughout this paper.

\section{Results}

Figure 1 shows on the same scale the $ICCF$s around the zero lag for all 42 objects. The $ICCF$ peak there
is typically the most prominent one, except for a few cases where additional (similar or even higher) maxima are present at 
significant distances from $\tau=0$ (e.g. PG 1012+008). For such peculiar cases, only the ``central'' maximum is used for the analysis.

\begin{table}
\caption{The quasar sample.}
\label{table:1}      
\begin{tabular}{lcccccc}
\noalign{\smallskip}
\hline\hline

Object 	& 	z & N & $\tau_{\rm obs}$  & $\log M_{\sun}$ & $\log \dot m$& $\tau_{\rm exp}$ \\

\hline

PG 0026+129	&	0.142	&	72	&	4	&	7.83	&	0.05	&	4.1	 \\
PG 0052+251	&	0.155	&	76	&	$-$4	&	8.75	&	$-$0.82	&	8.6	 \\
PG 0804+761	&	0.100	&	88	&	12	&	8.35	&	$-$0.30	&	7.0	 \\
PG 0838+770	&	0.131	&	29	&	30	&	7.99	&	$-$0.49	&	3.5	 \\
PG 0844+349	&	0.064	&	66	&	6	&	7.76	&	$-$0.48	&	2.5	 \\
PG 0923+201	&	0.190	&	25	&	16	&	9.09	&	$-$1.13	&	11.6	 \\
PG 0953+414	&	0.239	&	60	&	14	&	8.49	&	$-$0.20	&	9.4	 \\
PG 1001+054 &	0.161	&	26	&	$-$6	&	7.65	&	$-$0.02	&	2.9	 \\
PG 1012+008	&	0.185	&	23	&	$-$64	&	8.07	&	$-$0.32	&	4.5	 \\
PG 1048+342	&	0.167	&	31	&	8	&	8.24	&	$-$0.69	&	4.4	 \\
PG 1100+772	&	0.313	&	47	&	16	&	9.11	&	$-$0.75	&	16.0	 \\
PG 1114+445	&	0.144	&	25	&	2	&	8.42	&	$-$0.93	&	4.8	 \\
PG 1115+407	&	0.154	&	25	&	10	&	7.51	&	$-$0.14	&	2.2	 \\
PG 1121+422	&	0.234	&	26	&	$-$20	&	7.86	&	$-$0.23	&	3.5	 \\
PG 1151+117	&	0.176	&	23	&	28	&	8.44	&	$-$0.80	&	5.4	 \\
PG 1202+281	&	0.165	&	38	&	6	&	8.46	&	$-$1.05	&	4.7	 \\
PG 1211+143	&	0.085	&	24	&	16	&	7.83	&	0.05	&	4.1	 \\
PG 1226+023	&	0.158	&	45	&	$-$16	&	8.88	&	$-$0.01	&	19.6	 \\
PG 1229+204	&	0.064	&	45	&	0	&	8.00	&	$-$0.80	&	2.8	 \\
PG 1307+085	&	0.155	&	30	&	12	&	8.54	&	$-$0.65	&	7.2	 \\
PG 1309+355 &	0.184	&	32	&	$-$4	&	8.16	&	$-$0.42	&	4.7	 \\
PG 1322+659	&	0.168	&	28	&	0	&	8.08	&	$-$0.41	&	4.2	 \\
PG 1351+640	&	0.087	&	35	&	0	&	8.66	&	$-$1.06	&	6.3	 \\
PG 1354+213	&	0.300	&	26	&	2	&	9.46	&	$-$0.49	&	33.3	 \\
PG 1402+261	&	0.164	&	28	&	$-$2	&	7.85	&	0.02	&	4.1	 \\
PG 1404+226	&	0.098	&	28	&	$-$16	&	6.71	&	0.23	&	0.9	 \\
PG 1411+442 &	0.089	&	29	&	14	&	7.87	&	$-$0.54	&	2.8	 \\
PG 1415+451	&	0.114	&	30	&	2	&	7.80	&	$-$0.58	&	2.4	 \\
PG 1416$-$129&	0.086	&	25	&	6	&	8.92	&	$-$1.12	&	9.0	 \\
PG 1427+480	&	0.221	&	29	&	12	&	7.98	&	$-$0.34	&	3.8	 \\
PG 1444+407	&	0.267	&	27	&	$-$2	&	8.16	&	$-$0.12	&	6.0	 \\
PG 1512+370	&	0.371	&	36	&	$-$2	&	9.17	&	$-$0.87	&	15.9	 \\
PG 1519+226	&	0.137	&	33	&	26	&	7.78	&	$-$0.31	&	2.9	 \\
PG 1545+210	&	0.266	&	43	&	12	&	9.17	&	$-$0.93	&	15.1	 \\
PG 1613+658	&	0.129	&	64	&	0	&	8.95	&	$-$1.46	&	7.3	 \\
PG 1617+175	&	0.114	&	56	&	4	&	8.73	&	$-$0.88	&	8.0	 \\
PG 1626+554	&	0.133	&	30	&	0	&	8.37	&	$-$0.94	&	4.4	 \\
PG 1700+518	&	0.292	&	54	&	6	&	8.89	&	$-$0.45	&	14.3	 \\
PG 1704+608	&	0.371	&	40	&	40	&	9.20	&	$-$0.77	&	17.9	 \\
PG 2130+099	&	0.061	&	79	&	2	&	7.81	&	$-$0.37	&	2.9	 \\
PG 2233+134	&	0.325	&	33	&	$-$12	&	8.15	&	0.36	&	8.5	 \\
PG 2251+113 &	0.323	&	43	&	2	&	9.04	&	$-$0.62	&	15.7	 \\

\hline\hline
\end{tabular}
\end{table}

\begin{figure}
\centering
\resizebox{\hsize}{!}{\includegraphics{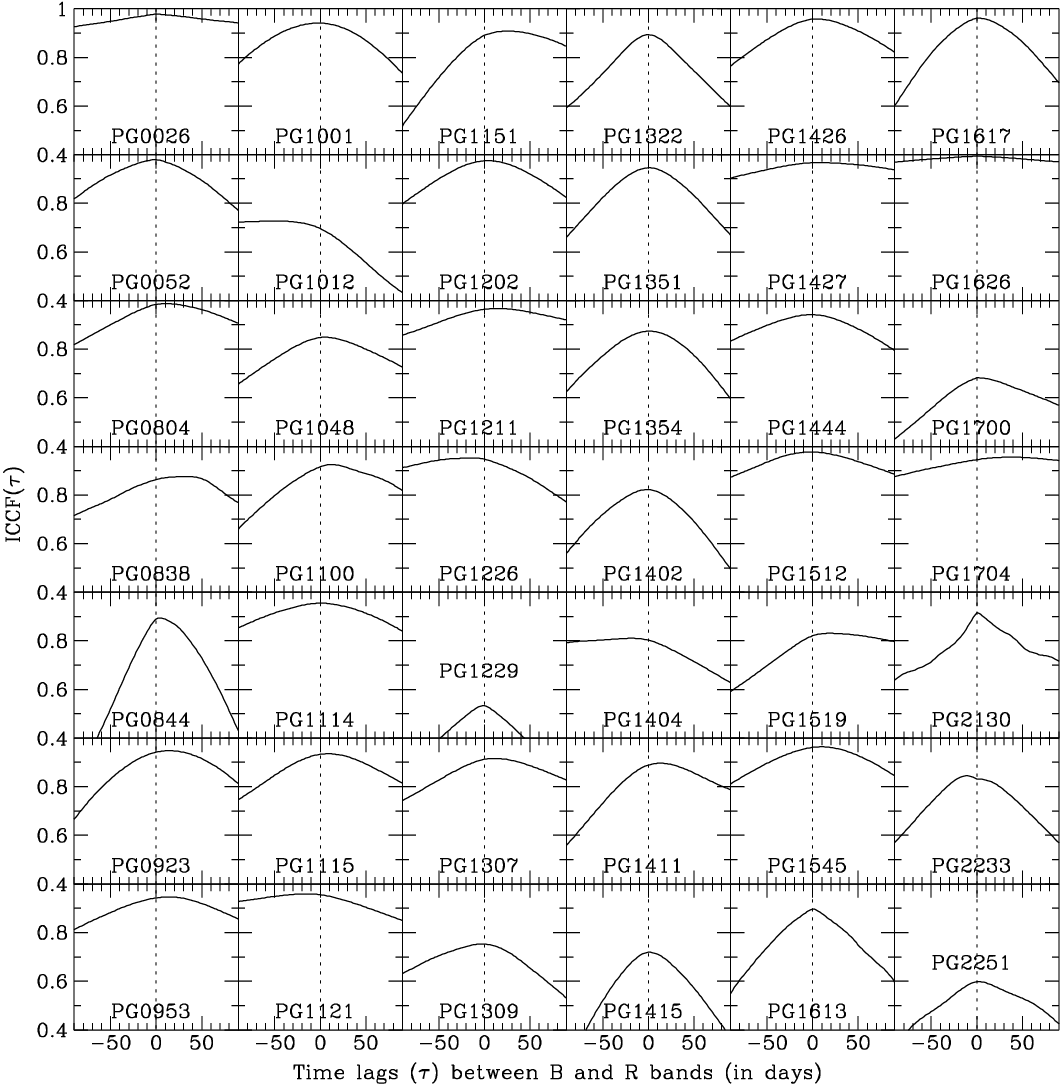}}
\caption{Interpolation cross-correlation functions of $B$ and $R$-band light curves as functions of the time delays (in days). A positive
time delay indicates $B$-band leading the $R$-band. The scale is the same for all objects.}
\label{Figure:1}
\end{figure}


Figure 2 presents the distribution of the $rest~frame$ time delays for all the objects from the sample. When all of them are included, the
average $\tau_{\rm rest}$ is +3.2 days ($\pm4.2$ at 95\% confidence interval) with a standard deviation of 13.2 days, and the 
median is +2.7 days. The $t$-test gives for the null hypothesis (assuming that the mean value for the parent population is zero) gives a 
$p$-value of 0.13, so the null hypothesis cannot be rejected at a significant enough level (meaning that the average time delay for the 
sample cannot be statistically distinguished from zero). However, the situation changes if one takes into account the presence of highly 
deviating delay values in the distribution (e.g. PG 1012+008 with $\tau_{\rm rest}\simeq-54$ days, Figure 2). Two more tests, based on the 
median value, reject the idea that the median delay is zero at least at 95\% level. First, the sign test, based on counting the number 
of values above and below the median, gives a $p$-value of 0.02. Second, very similar are the results from the signed rank test, which is 
based on comparing the average ranks of values above and below the median. These contradicting results from the mean and the median tests
indicate that omitting deviating values is perhaps justified. Once one such value is omitted (PG 1012+008, see above), the sample mean 
becomes +4.6 ($\pm3.2$ at 95\% conf. interval) days ($p=0.006$), and the median -- +3.5 ($p=$0.012 and 0.006 for the tests mentioned above, 
respectively). The remaining objects' distribution resembles Gaussian (shown in Figure 2), although the K-S test rejects the idea, 
strictly speaking.

\begin{figure}
\centering
\resizebox{6cm}{!}{\includegraphics{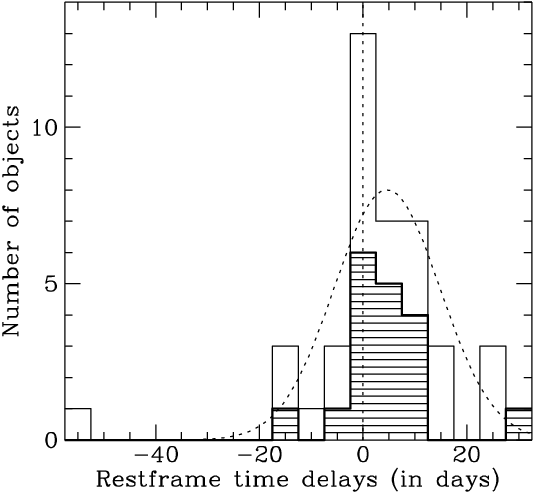}}
\caption{Distribution of the $rest~frame$ time delays ($ICCF$ peak positions) between $B$ and $R$-bands. A positive time lag means the blue-band 
light curve leads the red one. A Gaussian fit to the distribution is shown (one deviating object omitted). The shaded area shows the 
distribution of the well-sampled ($N>35$) objects only (see the text).}
\label{Figure:2}
\end{figure}


In a case that the reprocessing from a standard accretion disk is primarily responsible for the optical variations and 
respectively -- for the interband time lags, one would expect a wavelength dependent 
delay between the bands, which can be expressed (following Frank et al. 2002) as follows:

$\tau_{\rm B-R} \simeq 5 \dot m^{1/3}M_{\rm 8}(\lambda_{\rm R, 5000}^{4/3} - \lambda_{\rm B, 5000}^{4/3}) $ [days],

\noindent where $\dot m$ is the accretion rate in Eddington units, $M_{\rm 8}$ is the central mass in $10^{8}$ Solar masses, and
$\lambda_{\rm R, 5000}$, $\lambda_{\rm B, 5000}$ are the average wavelengths of the bands, measured in units of 5000\AA\
\footnote{This expression applies to the quasars' rest-frame. Due to the similar way the times and the wavelengths are affected 
by the red shift, for the observer's frame $\tau$ $reduces$ by a factor of $(1+z)^{1/3}$, which is only a few percent for 
this sample, and is much less than the expected errors.}. 
This expression is obtained under the assumption that the disk emits mostly due to a viscous heating and the
reprocessed radiation is only a small addition to the emitted flux. Thus, the disk rings will reprocess most effectively radiation 
of a wavelength close to the maximum of their Planck curves. Therefore, knowing the disk radial temperature distribution (Frank et al. 2002),
one can approximately assess the wavelength dependence of the time lag.
Another important assumption is that the reprocessed hard (X-ray) radiation comes from a location very close to the center. If this were not the
case, but the X-rays come from e.g. a jet base elevated high above the disk instead, an additional geometrical factor of $\sim\cos(\theta)$
$shortens$ the lag between $B$ and $R$-bands, due to the decrease of the path difference. Table 1 (the last column) presents the 
expected time lags, calculated based on the accretion parameters ($M$ and $\dot m$) from the previous two columns,
adopted from Kaspi et al. (2005).   

\begin{figure}
\centering
\resizebox{\hsize}{!}{\includegraphics{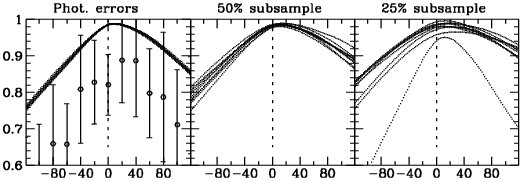}}
\caption{Various realizations of $ICCF$ as functions of the time delay (in days) for PG 0804+761, obtained by adding
appropriate random noise to the light curves (simulating the photometric errors) -- the left panel and using a random sub sample 
of the original sample (simulating scarcely observed cases) -- $\sim$50\% (the middle panel) and $\sim$25\% (the right panel). One sees 
that both effects, although influential, do not significantly alter the final result. For that particular case, the photometric 
error effect appears to be responsible for $\sim$0.4 days uncertainty in the $ICCF$ peak position, while the reduction of the original 
sample by a half -- for $\sim$1.3 days and by 75\% -- for $\sim$3.3 days.
In order to check the compatibility, the discrete cross-correlation method was also applied to the PG 0804+761 light curves 
(open symbols with error bars; the left panel). Here the peak cannot be defined so clearly, but the general shape of the curve is very similar. 
The slight vertical offset between the $ICCF$ and the $DCCF$ is perhaps a binning artifact (the time over which the $DCCF$ is averaged is 20 days)
and does not seem to affect the peak position.
}
\label{Figure:3}
\end{figure}

\section{Discussion}

\subsection{Sources of scatter}

\subsubsection{Photometric errors and sampling influence}

In order to get an idea about how much different uncertainties affect the time lags (the position of the $ICCF$ peak),
we performed two tests to the light curve data of a well sampled object, PG 0804+761, with a clearly defined positive lag, to
see if the photometric errors associated with the data points and the sampling can alter significantly the result. Figure 3 
(left panel) shows a number of $ICCF$s of the light curves with a random noise, appropriately added to mimic the expected photometric error. 
One sees that the effect is not significant, leading to a $\sigma_{\rm phot}\simeq0.4$ days uncertainty of the peak position. 
The sampling, as expected, can affect more significantly the peak position. This object is observed on $N=88$ epochs, 
while many others -- in less than 30. In order to study the effects of the scarce sampling, we randomly retrieved a sub sample of the 
original data points, which was further used for the interpolation and the cross-correlation analysis. The results for a number of 
simulations are shown in Figure 3, middle and right panels for 50\% and 25\% sub samples, respectively. One may get the impression of a significant 
scatter of the peak position, although it is due mostly to the different peak values (the maxima of the $ICCF$s are not normalized);
the peak position itself varies a little -- $\sim$1.3 days ($rms$) for the 50\%-th sub sample and $\sim$3.3 days for the 25\%-th one.   
Therefore, the photometric error and the sampling effects cannot alter significantly the lag results, at 
least for the well-sampled objects. Yet, for the very scarcely observed objects, the uncertainties of the peak position might be significant.
Based on the simulations described above, we found appropriate a very tentative error assessment approach, taking into account all sources of errors: 
$\sigma_{\rm \tau} \simeq \sqrt{\sigma_{\rm int}^2 + \sigma_{\rm phot}^{2} + 25(2-\log(N))^{2}}$, where we adopt $\sigma_{\rm int}=0.6$ (Sect. 3)
and $\sigma_{\rm phot}=0.4$ days. This expression may work reasonably well for our sample (where $N<100$), but should by no 
means be considered universal; see Gaskell \& Peterson (1987) for more details on the error issues. 

Note, that the statistical errors of the cross-correlation functions are very small due to the large 
number of interpolated magnitude values used and their effect on the peak position is negligible. 
If the observed time lags are as a result of reprocessing, the errors estimated above, following this simple approach, appear to be 
significantly underestimated for a number of reasons in comparison with the observed scatter (Fig. 4, see also Sect. 5.2).

\begin{figure}
\centering
\resizebox{\hsize}{!}{\includegraphics{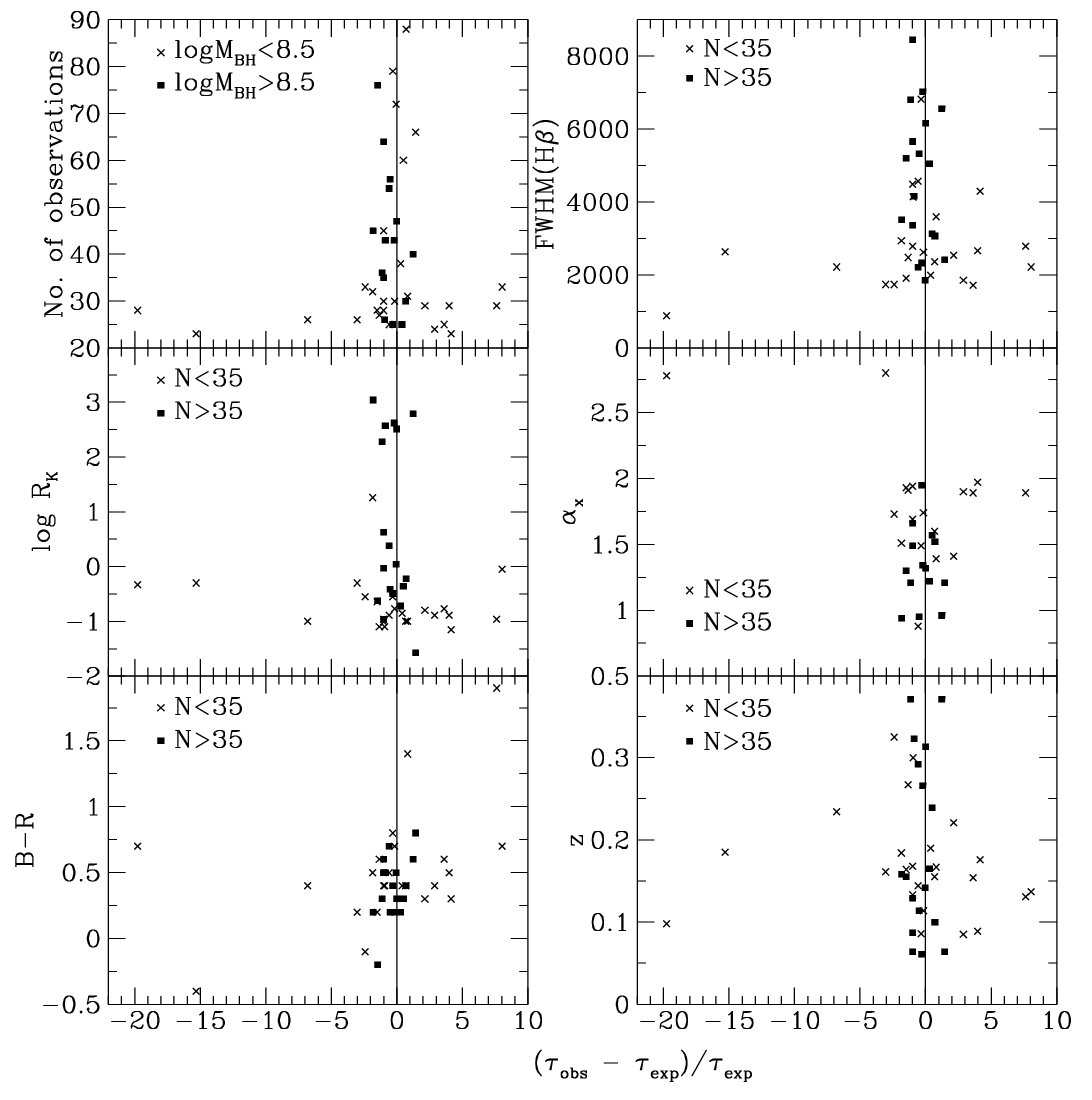}}
\caption{Normalized difference between the observed and the expected time lags $(\tau_{\rm obs}-\tau_{\rm exp})/\tau_{\rm exp}$ as functions 
of different quasar properties. The figure shows the influence of the number of the observations, $N$ (top-left panel, objects are separated by their 
black hole masses) and radio-loudness (Kellermann index, middle-left; (Kellerman et al. 1989); (B-R) color (bottom-left); 
H$\beta$ line width (in kms$^{-1}$, top-right); soft X-ray spectral index (middle-right); red shift (bottom-right) -- all they shown 
with different symbols, separated by the number of the observations. As the top-left panel shows, the scatter between the 
observed and the expected times is significant only for $N<35$, so this value is further used in the remaining panels as a separation criterion 
between the well-sampled and the under-sampled objects.
}

\label{Figure:4}
\end{figure}

\subsubsection{Other sources of scatter}

Although the observed and the calculated lags appear to be broadly consistent for the well-sampled objects (Sect. 5.2), the scatter is 
significant. Except for the errors, described in Sect. 5.1.1, there are also several other factors that may contribute to the scatter:

\begin{itemize}

\item \textbf{$ICCF$ method limitations}. It is not clear to what extent replacing large missing parts of the real light curve with straight 
lines, randomly or systematically alters the lag results. Note, that the errors discussed in Sect. 5.1.1 are not the errors 
introduced by the interpolation itself.

\item \textbf{Correlated photometric errors}. If the photometric errors of $B$ and $R$-bands happen to be correlated to some extent 
(due to variable seeing, atmospheric transparency changes, reduction errors, low-level variability of comparison stars, etc. common problems), 
leading to a common offset of both $B$ and $R$-magnitudes for a given epoch of observation, the time lag between the bands will 
naturally decrease (as an absolute value), as these effects will lead to an increased correlation between the data sequences at zero lag.

\item \textbf{Accretion parameters' uncertainties}. Uncertainties in the black hole mass and accretion rate (possibly systematic!) 
naturally increase the scatter.

\item \textbf{Central source location}. The height of the central irradiating source above the disk is another unknown that may well vary from object to
object, as well as in time (Sect. 5.2). Furthermore, nothing guarantees that this source is located along the central axis; it can simply be an
active region, located somewhere above the disk, even closer to the R-band emitting parts than to the B-band such (Sect. 1).

\item \textbf{Emission lines' contribution}. The broad emission lines (H$\alpha$ and H$\beta$) can also contribute to the scatter, as they 
may fall into the $B$ and $R$-bands and are expected to vary with a lag behind the continuum. For most of the objects, however, the broad-line 
response times are significant, typically $\sim$100 days, which is much longer than the average continuum lags. Also, for most of the objects, 
the red shift is high enough to move H$\alpha$ out of the $R$-band, but around $z\simeq0.2$ H$\beta$ enters this band.

\end{itemize}

\subsection{Quasar properties}

Except for the uncertainties of different nature (see above), the scatter between the expected and the observed lags can also
be attributed to the possibility that the reprocessing may not be the primary driver of the optical variability for some of the
objects. As mentioned before, a weak blazar component can also play a role, in addition to other (possibly unknown) mechanisms.
If so, one may expect to be able to differentiate between the quasars, for which the reprocessing is responsible for the observed lags
from those, for which this mechanism is different, based on other quasar properties, such as radio loudness, 
X-ray spectral index, continuum colors, optical spectra, etc. The idea is that these different mechanisms may leave their signatures on the  
observed quasar appearance. We tested the normalized time lag difference against various quasar observables, including luminosity, accretion rate, 
equivalent widths of H$\beta$ and [OIII] lines and their ratio, FeII/H$\beta$ ratio, radio power, X-ray to optical continuum 
index\footnote{Data are taken from Giveon et al. (1999) and Kaspi et al. (2005)}, as well as CIV$\lambda$1549 equivalent width and shift\footnote{
Data are taken from Bachev et al. (2004) and Sulentic et al. (2007)}, but found no significant correlations, except for some tendencies. 
The most interesting cases are shown in Figure 4. The top-left panel shows the influence of the number of observational 
epochs -- one sees that for $N \ga 35$ the scatter reduces significantly, meaning that for many objects, the number of the observations 
may be simply not high enough to reveal the true time lags. Based on this observation, all the remaining panels of this figure show
separately the objects for which $N>35$ (as filled squares) from the remaining, undersampled objects (as crosses).
Interestingly, this tendency is not that strong for the higher-mass objects ($\log M_{\rm BH}>8.5$), for which the scatter appears to be 
generally small for all $N$ (only in the top-left panel the separation is based on $\log M_{\rm BH}$).
The radio-loudness does not seem to play a significant role in the scatter (middle-left), meaning probably that a possible 
blazar component does not contribute significantly to the optical variability for this sample.
A weak tendency for the bluer object to have a delay between the bands shorter than expected, even for the well-sampled objects, is shown 
in the bottom-left panel (Figure 4). It is not clear how to explain this effect, if real at all, but it may be connected to the 
way continuum is generated (i.e. the exact accretion disk structure and properties, being perhaps different from the standard model).
The top-right panel shows the influence of the of the width of the broad lines (H$\beta$). The scatter appears to be most significant
for the narrower-line (FWHM$\la$3000 kms$^{-1}$) objects, which may have implications for possible quasar population differences 
(e.g. Sulentic et al. 2000). The soft X-ray spectral index panel (middle-right) reveals an interesting possibility that the optical variations of 
the harder ($\alpha_{\rm X}\ga 1.8$) objects are perhaps less likely to be attributed to reprocessing in a $standard$ accretion disk, 
which may have implications for the origin of the soft X-ray excess.  
Finally, the bottom-right panel demonstrates that, indeed, the emission lines around $z\simeq0.2$ may also play a role in the continuum 
time lags, as the scatter appears to be the largest there (see Sect. 5.1.2 for more details). 

If we take into account only the best sampled objects ($N \ga 35$), the average normalized difference between the observed 
and the expected time lags, $\overline{(\tau_{\rm obs}-\tau_{\rm exp})/\tau_{\rm exp}}$, is $-0.32$, slightly less than the expected
value of zero. Should there be no systematic errors involved in the estimation of $\tau_{\rm exp}$, and in a case that the reprocessing 
is mainly responsible for the lags, another possibility emerges, i.e. the irradiating source is located above the disk, as
mentioned previously. The value of $-0.32$ corresponds roughly to an elevation angle of $\simeq45$ deg., meaning a height above the disk
in order of the distances at which the optical continuum is generated, i.e. 100 -- 1000 gravitational radii (Ar\'evalo et al. 2008), 
raising the possibility that the jet base is the primary source of irradiating X-rays (see however Czerny \& Janiuk 2007, for a warm absorber
interpretation). 
This interpretation, however, due to the large scatter, cannot be justified statistically with the available optical data, so it has to 
be considered only as an interesting possibility.

\section{Conclusions}

In this paper we analyze the light curves and study the time lags between the optical continuum bands for a large sample 
of quasars. In spite of the significant scatter, we show that the lags are broadly consistent with the reprocessing model, 
according to which the optical variations are largely due to the reprocessing of the central X-ray radiation in a
surrounding thin accretion disk. There are also some indications that the central X-ray irradiating source may be located at some
height (a few hundred Schwarzschild radii) above the disk plane, representing perhaps a (failed) jet base.
The deviations of the observed lags from the expectations, assuming the reprocessing model, do not seem to correlate 
significantly with any other quasar properties and are probably due mostly to the scarce sampling. 
The paper also demonstrates that the broad-band optical monitoring of quasars could be a powerful tool to study the central engines,
provided the light curve is well sampled. A small robotic telescope (e.g. 0.4--0.6m), dedicated to monitoring a large sample of
brighter quasars once every 2--3 days, with an accuracy of 0.02 mag or better in 3 filters, might be able to clarify the role of
reprocessing in quasar variability.


\end{document}